\documentclass[]{spie}  

\usepackage{amsmath,amsfonts,amssymb}
\usepackage{graphicx}
\usepackage[colorlinks=true, allcolors=blue]{hyperref}
\usepackage{setspace}
\usepackage{tocloft}
\usepackage{xcolor}
\usepackage{xspace}
\usepackage[perpage]{footmisc}

\title{Observation Scheduling and Automatic Data Reduction for the Antarctic telescope, ASTEP+}

\author[a]{Georgina Dransfield*}
\author[b]{Djamel M\'ekarnia}
\author[a]{Amaury H.M.J. Triaud}
\author[b]{Tristan Guillot}
\author[b]{Lyu Abe}
\author[c]{Lionel J. Garcia}
\author[c]{Mathilde Timmermans}
\author[d,e]{Nicolas Crouzet$^{\star,}$}
\author[b]{Fran\c cois-Xavier Schmider}
\author[b]{Abdelkrim Agabi}
\author[b]{Olga Suarez}
\author[b]{Philippe Bendjoya}
\author[e]{Maximilian N. G\"unther$^{*,}$}
\author[b]{Olivier Lai}
\author[e]{Bruno Merín}
\author[b]{Philippe Stee}
\affil[a]{School of Physics \& Astronomy, University of Birmingham, Edgbaston, B15 2TT, Birmingham, UK}
\affil[b]{Universit\'e C\^ote d'Azur, Observatoire de la C\^ote d'Azur, CNRS, Lagrange Laboratory, Nice, France}
\affil[c]{Astrobiology Research Unit, Université de Liège, 19C Allée du 6 Août, 4000 Liège, Belgium}
\affil[d]{Leiden Observatory, Leiden University, Postbus 9513, 2300 RA Leiden, The Netherlands}
\affil[e]{European Space Agency (ESA), European Space Research and Technology Centre (ESTEC), Keplerlaan 1, 2201 AZ Noordwijk, The Netherlands}

\authorinfo{Further author information: (Send correspondence to G. Dransfield)\\
G Dransfield: E-mail: gxg831@bham.ac.uk\\
$^\star$ ESA Research Fellow (2018-2021) \\
$^*$ ESA Research Fellow}

\newcommand{\prose}{\texttt{prose}}
\newcommand{\astep}{{{ASTEP}}\xspace}
\newcommand{\tess}{{\it{TESS}}\xspace}
\cftpagenumbersoff{figure}
\cftpagenumbersoff{table} 
 
\begin{document} 
\maketitle

\begin{abstract}
The possibility to observe transiting exoplanets from Dome C in Antarctica provides immense benefits: stable weather conditions, limited atmospheric turbulence, and a night that lasts almost three months due to the austral winter. However, this site also presents significant limitations, such as limited access for maintenance and internet speeds of only a few KB/s. This latter factor means that the approximately 6\,TB of data collected annually must be processed on site automatically, with only final data products being sent once a day to Europe. In this context, we present the current state of operations of ASTEP+, a 40\,cm optical telescope located at Concordia Station in Antarctica. Following a successful summer campaign, ASTEP+ has begun the 2022 observing season with a brand-new two-colour photometer with increased sensitivity. A new \texttt{Python} data analysis pipeline installed on a dedicated server in Concordia will significantly improve the precision of the extracted photometry, enabling us to get higher signal-to-noise transit detections. The new pipeline additionally incorporates automatic transit modelling to reduce the amount of manual post-processing required. It also handles the automatic daily transfer of the photometric lightcurves and control data to Europe. 
Additionally, we present the \texttt{Python} and web-based systems used for selection and scheduling of transit observations; these systems have wide applicability for the scheduling of other astronomical observations with strong time constraints. We also review the type of science that ASTEP+ will be conducting and analyse how unique ASTEP+ is to exoplanet transit research.
\end{abstract}


\section{Introduction}
\label{sec:intro}  

ASTEP (Antarctic Search for Transiting ExoPlanets) was initially conceived in 2006 as a photometric telescope to search for transiting exoplanets in dense fields of stars \cite{Fressin+2007}. It was installed at the Concordia Station, located on Dome C in Antarctica, at the end of 2009, and successfully began its operations in 2010 \cite{Daban+2010, Guillot+2015}. The telescope remained on site until 2013 exploiting the near-continuous winter polar nights and excellent weather conditions of the site \cite{Crouzet+2010, Crouzet+2018}. It led to the first ground-based observation of a secondary eclipse of an exoplanet in the visible \cite{Abe+2013} and the discovery of tens of transiting planet candidates \cite{Mekarnia+2016}. At that time, internet connection with the Concordia station was extremely limited, requiring the full-time presence of an astronomer on the site. Observations were saved on hard drives and fully analyzed in Europe the year after. 

The return of ASTEP at the end of 2016 for a continuous observation of $\beta$~Pictoris \cite{Mekarnia+2017, Lagrange+2019, Kenworthy+2021A&A} required a different strategy, i.e., an automatic processing of the lightcurves and their transmission to Europe. Improvements in the internet connection with the Concordia station implied that an automatic transmission of limited amounts of data (of order $10~\rm MB/day$) was possible, using the HERMES system implemented by PNRA ({\it Programma Nazionale di Ricerche in Antartide}). A new server, adapted to the automatic processing of the $\sim 6$\,TB of data per season, was sent to Concordia with a dedicated \texttt{IDL} pipeline, based on aperture photometry\cite{Mekarnia+2017}. With the successful launch of \tess (Transiting Exoplanet Survey Satellite)\cite{Ricker+2015}, the ASTEP observation program naturally transitioned to a follow-up of transiting exoplanet candidates, focusing on those with long orbital periods (10-100+ days) and contributing to many discoveries\cite{Bouma+2020AJ, Dawson+2021, Dong+2021, Grieves+2021, Burt+2021, Kaye+2022, Wilson+2022, Mann+2022, Christian+2022, Dransfield+2022}. 

The geographic location of ASTEP is extremely interesting in complement to several key exoplanetary space missions. Of course, the Antarctic polar night is highly favorable for an efficient observation of transiting planets, in particular for those with long orbital periods and for those with long transit durations (the two being correlated). The combination of observations from mid-latitude sites in e.g., Chile, and Antarctica is particularly efficient\cite{Fruth+2014}. However, the main asset is probably the fact that targets in the southern continuous viewing zones of the \tess, \textit{JWST} and \textit{Ariel} space missions are circumpolar and can be observed by ASTEP all the time. In counterpart, access to Antarctica is limited to about 3 months, between November and February, only a low-bandwidth internet connection is available, and the instrument must cope with temperatures ranging from about $-10^\circ$C to $-80^\circ$C. This makes operations with ASTEP intermediate between ground-based telescopes and space-based missions.

Starting in 2022, thanks to support from the University of Birmingham, ESA, INSU, and the Laboratoire Lagrange, a new camera box\cite{Crouzet+2020} could be installed, starting ASTEP+ observations in two colors. This required a new pipeline, but also an adaptation of the scheduling system to cope with the large number of potentially interesting targets to be observed.

\textcolor{black}{In this paper we will describe the current status of ASTEP+. First we show how different ASTEP+'s sensitivity to transiting planets is to more traditional observatories in Section~\ref{sec:long}. In Section \ref{sec:ObsSched} we outline our current scheduling needs and how they are shaped by our team's science goals. We also describe the systems we have designed to meet these needs and our plans for further development. In Section \ref{sec:pipeline} we provide a detailed description of our new automatic \texttt{Python} data analysis pipeline, as well as a quantitative comparison with the system it replaces. In Section \ref{sec:future} we map out in brief the possible future directions of the ASTEP+ project, and we conclude in Section \ref{sec:conclusions}.}

\section{Sensitivity to long period planets and long duration transits}\label{sec:long}

\textcolor{black}{Prior to the start of the 2020 observing season, ASTEP had been dedicating its exquisite resources first to a blind exoplanet survey of its own \cite{Mekarnia+2016}, and then to observations of the $\delta$-Scuti pulsator $\beta$~Pictoris \cite{Mekarnia+2017}. In early 2020, ASTEP joined TFOP (\tess Follow-Up Observing Programs) Sub-Group 1 (SG1 hereafter) to support photometric follow-up and confirmation of \tess planet candidates.}

\textcolor{black}{In the context of \tess follow-up, \astep has a lot to offer; in particular, \astep can truly excel when it comes to observing long transits in full, as well as the transits of long-period planets. The current list of TOIs (\tess Objects of Interest) has 5767 targets, of which 3053 are observable from the South\footnote{Taken from \url{https://exofop.ipac.caltech.edu/tess/view_toi.php}, correct on the 22th of June 2022.}. Of these, 285 have transits lasting at least five hours, and 186 have periods of at least 20 days. These 471 candidate planets fall perfectly in \astep's niche. }

\textcolor{black}{In order to quantitatively compare \astep's sensitivity to long transit/long period candidates, we created a synthetic sample of 2500 planet candidates. Each was given a transit epoch ($\rm t_0$) drawn randomly from \tess's first year, a period of between $\rm 20-200\,days$, and a transit duration between $\rm 5-15\,hours$. To keep the on-sky distribution of the sample as realistic as possible, we assigned to each a set of coordinates (right ascension, declination) of real TOIs as taken from the current TOI list described above. Using this sample, we simulated observability of full transits by \astep and three other SG1 observatories: SPECULOOS-South (SSO) \cite{Delrez2018} in Chile, South African Astronomical Observatory (SAAO)\cite{saao2013} in South Africa, and Hazelwood Observatory\cite{hazelwood2019} in Australia. The simulation was carried out using the \texttt{Python} package \texttt{astroplan}\cite{2018AJ....155..128M}, and observability was tested for a five-year period starting from the 01 January 2022.}

\begin{figure}
    \centering
    \includegraphics[width=0.9\textwidth]{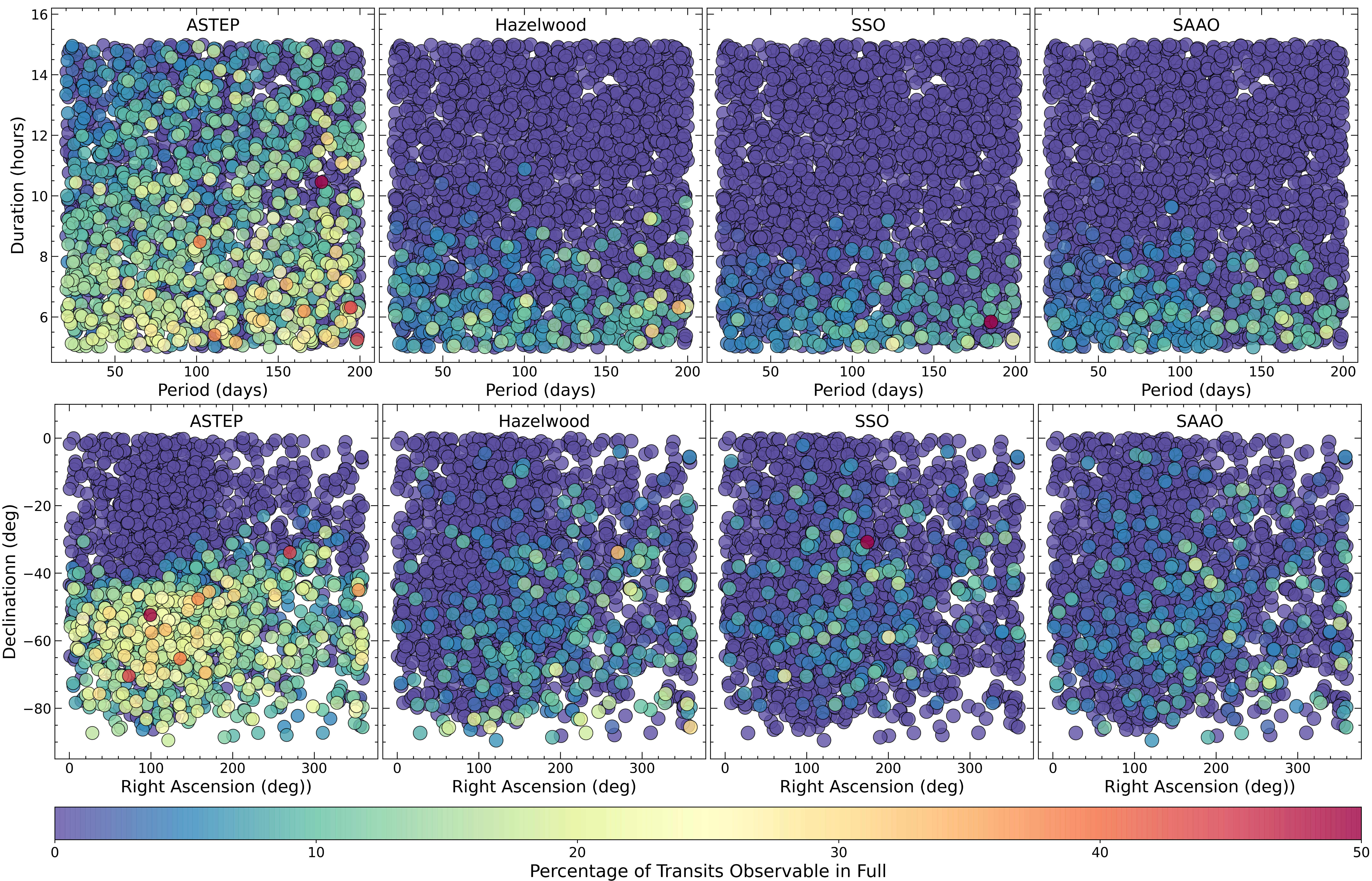}
    \caption{Results of the observability simulation for our synthetic sample of 2500 long period/long duration transiting planets. In all panels points are coloured according to the percentage of transits observable in full. Top row: duration and period distribution of the sample. Bottom row: the distribution of the sample on the sky.}
    \label{fig:obs_sim}
\end{figure}

\textcolor{black}{The results of our simulation are presented in Fig. \ref{fig:obs_sim}, with the top row of panels showing the distribution of planets according to period and duration and the bottom panel showing the planets' distribution on the sky. There are two very clear conclusions to be drawn from the top panels; firstly, we can see that \astep is the only observatory that can \textit{consistently} observe full events for both infrequent and long duration transits. The other observatories have almost no observability above eight hours, while \astep maintains in excess of 20\% for some transits lasting over 12 hours. Secondly, we can see that while all observatories can observe \textit{some} long-period transits in full, their percentage observability is lower than \astep's, and where long periods intersect with long transits, there are no transits observable. We find that on average \astep is more than 8 times more likely to observe a full transit for long period planets, and for the duration bins where other telescopes have non-zero observability, \astep is almost 40 times more likely to catch a full event.}

\textcolor{black}{In the lower panels we can see how the sample of synthetic planets is spread on the sky, and we note that the bulk of \astep's zero-observability objects are concentrated in an area of the sky that is completely invisible to us. Circumpolar and very southern targets are high in the sky all year long, while more northern targets (above declinations of $-40^\circ$) rise during the austral summer and set before \astep's observing season begins. }

\begin{figure}
    \centering
    \includegraphics[width=0.9\textwidth]{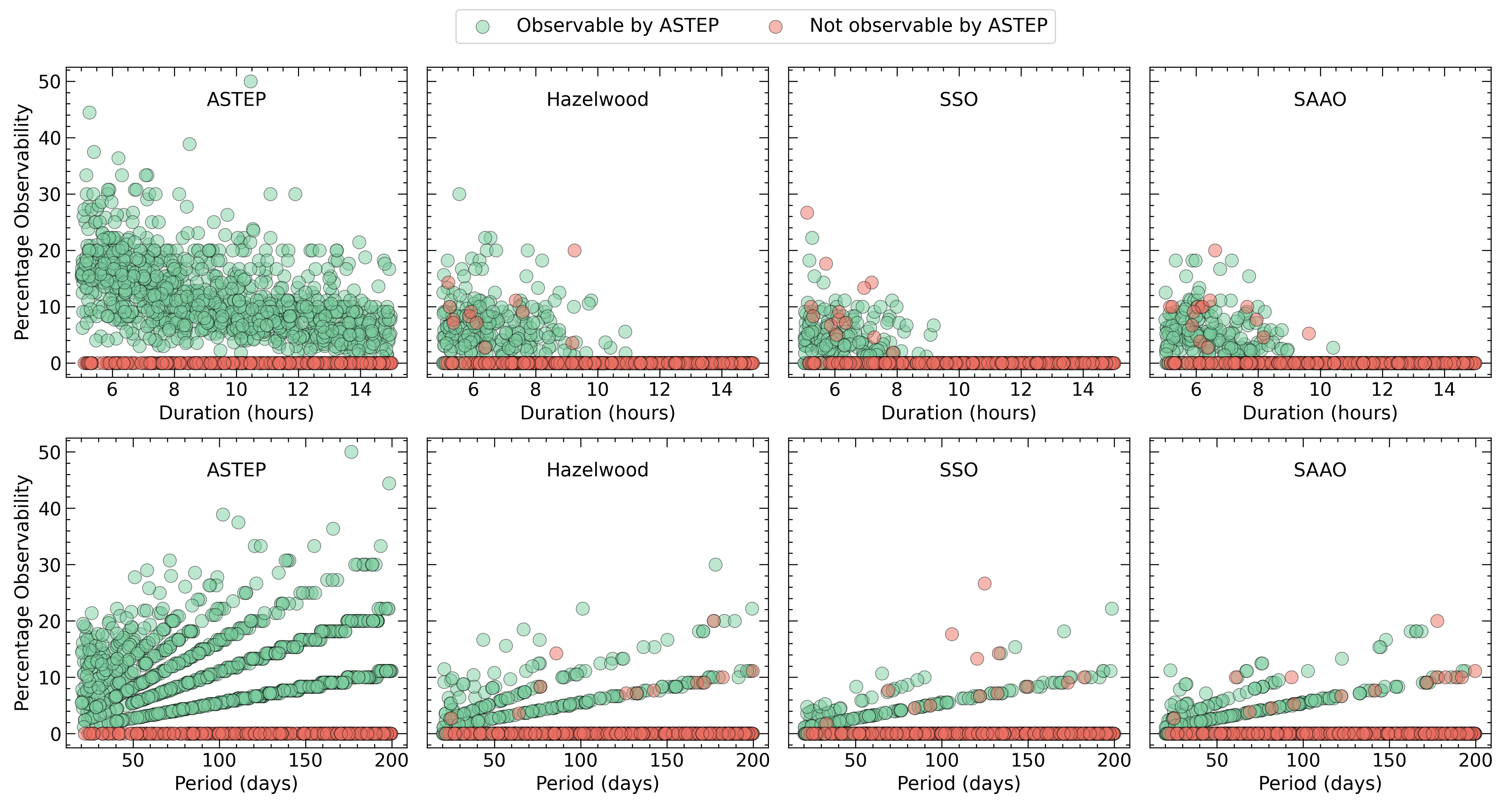}
    \caption{Which systems can never be observed? Having removed all systems north of $-40^{\circ}$, we show the systems that ASTEP still cannot observe to check if they are observable by other telescopes. Green points in all panels are systems where ASTEP can observe at least one transit in full, while red points have zero transits observable by ASTEP in the next five years. Top row: observability in terms of duration; bottom row: observability and period distribution. }
    \label{fig:comparison}
\end{figure}

\textcolor{black}{In Fig. \ref{fig:comparison} we filter out all objects north of $-40^{\circ}$ and compare observability of the objects still invisible to \astep. In all panels, objects that \astep can observe at least one full transit for are plotted in green, while planets with zero observability are plotted as red circles. Crucially, we note that for all systems where \astep cannot observe a single transit in the next five years, only 2.3\% can be observed by a different observatory.}

\section{Observation Scheduling}
\label{sec:ObsSched}



The scheduling of our observations is heavily influenced by the type of observations we perform. At the moment, the main task of ASTEP+ is to confirm planetary candidates identified by ExoFOP (Exoplanet Follow-up Observing Programme), an international follow-up collaboration. Our lightcurves of transiting planets are necessary to validate \tess's planetary candidates for the following reasons:\vspace{-0.5em}
\begin{itemize}
    \item \tess has a PSF (Point Spread Function) of about 30 arcseconds \cite{Ricker+2015}, meaning that many transit events are in fact deep eclipse produced by faint background objects diluted by a foreground bright star (which is then usually identified as a transiting planet candidate by the \tess automated pipelines);\vspace{-0.5em}
    \item Sometimes,  transit events identified by \tess are false alarms, caused by instrumental systematics or an unlucky coincidence of stellar activity;\vspace{-0.5em}
    \item We can check odd and even transit events and verify their depth are consistent with one another\cite{Dransfield+2022}, as a way to rule out blended eclipsing binaries (in non visually resolved binaries);\vspace{-0.5em}
    \item We measure the transit depth and compare it to \tess's and to other photometric bands in order to perform a {\it chromaticity check}\cite{Dransfield+2022}, another way to rule out blended eclipsing binaries, which works if the secondary component has a significantly different $T_{\rm eff}$ from the primary. For most of its operations ASTEP operated with a single filter, but recently we installed two cameras separated by a dichroic, allowing us to perform chromaticity check directly (see companion paper by Schmider et al.);\vspace{-0.5em}
    \item We verify the transit ephemerides, notably the orbital period and time of transit;
\end{itemize}  

Exoplanet transits are short compared to their orbital periods, and happen at specific times. This means that we need a scheduler able to handle those. In addition, new systems are identified by \tess with regularity, and some are removed (e.g. when we demonstrate the system is likely not a planet). Again here this forces our scheduler to adapt to new situations on a weekly basis.

While every night can be filled with observations of \tess Objects of Interests (TOIs), we remain scientists with interests in specific types of planets. This places further restrictions on the type of scheduler. First we prioritise systems aligned with our science goals (be they confirmed TOIs, non-confirmed TOIs, or unrelated systems altogether). Because this rarely fills the schedule, we also select TOIs opportunistically but also with a prioritisation that aligns with our preferences (see Section \ref{sec:long}.)

In addition, we also perform filler observations for a few hours a night to create long timeseries. Typically this involves observing $\beta$~Pictoris to search for a transit of the Hill sphere of its directly imaged planets\cite{Kenworthy+2021}, monitoring J0600, a system with a candidate circumplanetary ring transiting a star (Kenworthy et al. in prep), and now some flaring stars.

Finally, our team is international, with partners in France, the United Kingdom and the Netherlands. We are not able to meet physically to decide targets, therefore requiring a web-based tool.

\subsection{Early solutions}

\begin{figure*}[t!]
    \centering
    \includegraphics[width=0.5\textwidth]{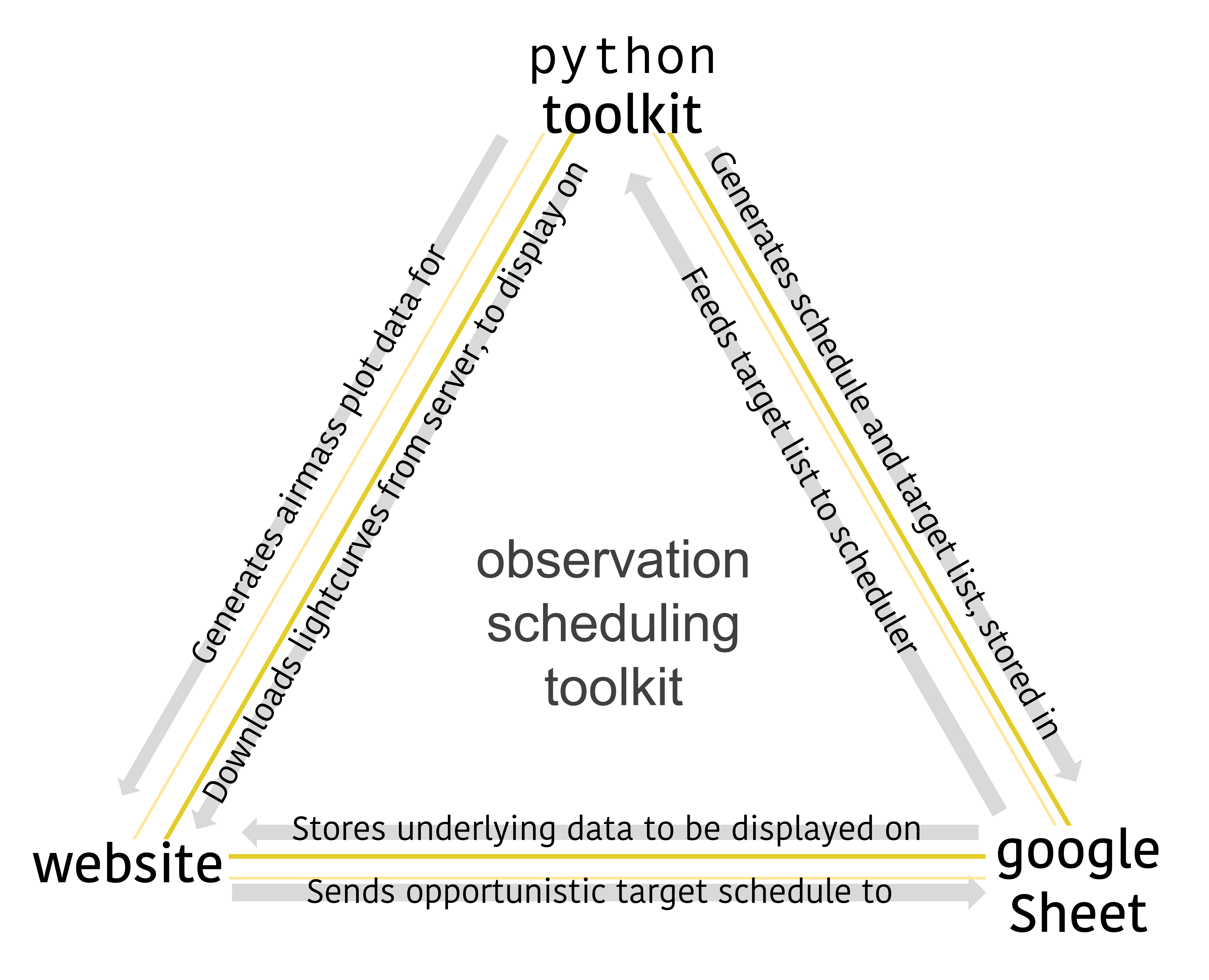}
    \caption{Schematic of the ASTEP+ scheduling toolkit, showing how the three elements are interconnected. The \texttt{Python} toolkit generates data, the Google Sheet holds the data, and the website both displays and updates the data. }
    \label{fig:toolkit}
\end{figure*}

\textcolor{black}{Upon joining ExoFOP at the start of 2020, ASTEP gained access to the \texttt{\tess Transit Finder}\cite{TTF} (\texttt{TTF} hereafter), a web-based tool that allows observers to check for upcoming transits of targets of interest. It can also be used to search for all transits occurring in a given date range, which allowed the team to get a feel for the different categories of transiting candidates \tess was producing.}

\textcolor{black}{Throughout ASTEP's first observing season with \tess, the \texttt{TTF} output was saved to \texttt{.csv} files and then scored using an \texttt{IDL} routine according to how well they aligned with our niche (Crouzet et al. in prep). Airmass plots were then generated to highlight when in the night the transits would happen, and the information was saved to \texttt{.pdf} files which were then circulated to the team in advance of our weekly target selection meetings. Targets selected for observation each night were recorded in a Google Doc\cite{gsuite} which was shared with the team.}

\textcolor{black}{Both elements of this initial system were limited. The generation of the \texttt{.pdf} files required that each team member check the Google Drive for new files on a weekly basis. Our use of a Google Doc to record the upcoming schedule also rapidly became time consuming; additionally, the important information contained therein was vulnerable as the whole team had access. Finally, manual input of information meant that small inaccuracies could cause us to miss important observations.}

\textcolor{black}{The earliest incarnations of the system we use now harnessed the G-Suite\cite{gsuite} \texttt{Python} APIs (Application Programming Interfaces) to read and write information to a Google Sheet instead of a Google Doc. G-Suite provides excellent collaborative working tools for international teams such as ours, but limiting vulnerability of underlying data while ensuring everyone could access crucial information became a top priority. We also found that time could be saved generating the airmass plots by making use of the \texttt{TTF} API rather than manually downloading the output.}

\textcolor{black}{One final limitation we encountered was that most our transit observations were opportunistic. Moving forward, the ambition was to define cohesive observing programs of scientific interest to the team.}



\subsection{Current Systems}

\textcolor{black}{The system we have implemented since the start of the 2021 observing season consists of three components: a Google Sheet;  a \texttt{Python} toolkit; and a website. Each of these elements and their key functions are described in the sections that follow. The interconnections between all three elements of the toolkit are summarised in Fig.~\ref{fig:toolkit}.}



\subsubsection{Generating the data: \texttt{Python} toolkit}
\label{sec:tools}

\begin{figure*}[t!]
    \centering
    \includegraphics[width=0.8\textwidth]{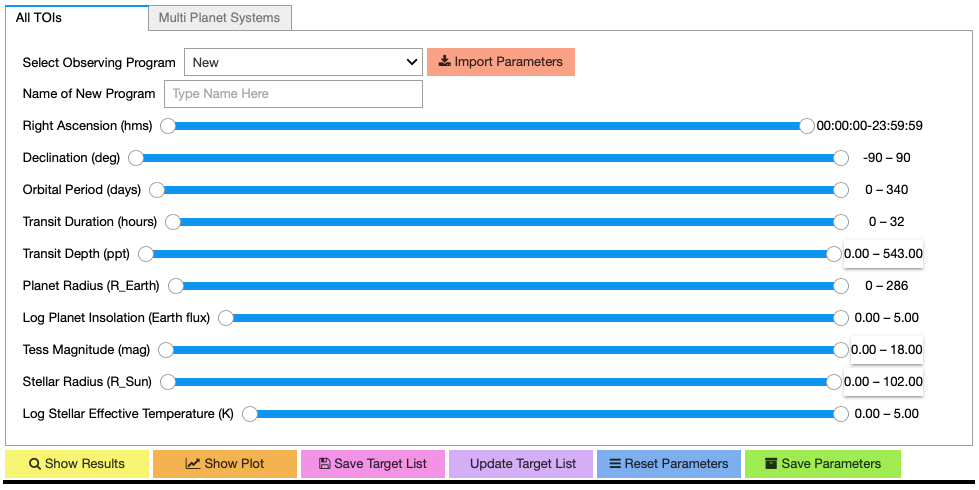}
    \caption{Screenshot of the graphical user interface for the target selection toolkit.}
    \label{fig:target_lists}
\end{figure*}

\textcolor{black}{As described above, one of ASTEP's key goals that emerged from the 2020 observing season was to define a list of targets that were of particular scientific interest to us. With an ever-growing list of candidates emerging on a monthly basis from \textit{TESS}, there was a need for a simple way to filter through the list of TOIs (\textit{TESS} Objects of Interest) to find those best suited to our unique context. This motivated the development of our \texttt{TESS\_target\_list} tool, written in \texttt{Python} and used via a \texttt{Jupyter Notebook} with an \texttt{ipywidgets} GUI. This simple tool makes use of intuitive sliders and dropdown lists to make target lists based on user-selected parameters. We present a screenshot of the \texttt{TESS\_target\_list} tool in Fig.~\ref{fig:target_lists}.}

\textcolor{black}{With pre-selected target lists now in place, a vast amount of scheduling can be done ahead of time. Computing the timings of upcoming transits using linear ephemerides is trivial, however the \texttt{Python} package \texttt{astroplan} \cite{2018AJ....155..128M} does this with elegance and simplicity. Not only can it compute the timings of upcoming events for eclipsing systems, both stellar and planetary in nature, but it will also output observability of said events for different observatories.}

\textcolor{black}{Leveraging both \texttt{astroplan} and the Google Sheets \texttt{Python} API, we developed the scheduler: \texttt{schedule\_ahead}. This software reads the target list directly from a `Master Target List' located on a Google Sheet (see Fig.~\ref{fig:toolkit}) and imports the stellar coordinates and orbital ephemerides for each of the targets, provided they have an `Observe' status of \textit{`Active'}. User inputs are an observatory (\astep, in this case), a start date, and a number of days to schedule for; these are combined with the system parameters from the target list and fed into \texttt{astroplan}. Only events where at least one of ingress and egress are observable are then output. The user then has the option of pushing the outputs to the `Schedule' Google Sheet and the `Web Schedule'; the latter will be described in Section \ref{sec:website}. }

\textcolor{black}{Before running the scheduler each week, there is the option to update the transit ephemerides of candidates in our target list by checking the \texttt{TTF}. This ensures that refinements resulting from other teams' observations are included, but crucially it tells us if a candidate has been retired as a false positive, i.e. it has been conclusively shown not to be a planet.}

\textcolor{black}{Our nights are very seldom filled perfectly with pre-scheduled events; in fact there are often multiple events overlapping in one night and a decision has to be made about what should be observed (see Section \ref{sec:website}). Additionally, the schedule often has large gaps (see Fig.~\ref{fig:web_sched}) that can be filled with our pre-selected filler targets or opportunistic observations of transits. In order to facilitate the latter, we use a custom \texttt{Python} script to search the \texttt{\tess Transit Finder}\cite{TTF} for transits of \tess candidates within a given date range. The results of the search are then scored according to how well each target suits our scientific interests, and airmass plot data are computed for the top 30 events using \texttt{astropy}\cite{astropy:2013,astropy:2018}. These data, along with system parameters of interest, are stored in the Google Sheet. This script therefore completely replaces the \texttt{IDL} script which generated the \texttt{.pdf} files.}

\subsubsection{Holding the data: Google Sheet}
\label{sec:gsheet}

\textcolor{black}{The team's Google Sheet holds all the underlying data generated by the \texttt{Python} toolkit, and updated and read by the website, as shown in Fig.~\ref{fig:toolkit}: the target list, the schedule, the airmass plot data, and the team's thoughts. }

\textcolor{black}{The Master Target List contains host star parameters, transit ephemerides and \texttt{TTF} meta-data (such as observing notes) for each target we have chosen to observe. The data for this sheet are generated by our \texttt{TESS\_target\_list} tool, and updated by the ephemerides checking script before running the scheduler each week. Targets are grouped into observing programs and are each given an `Observe' status: \textit{`Active'}, \textit{`Standby'} or \textit{`Retired'}. This flag is read by the scheduler and can be updated manually or programmatically.}

\textcolor{black}{The observation schedule is saved in two worksheets; one is human readable with observation and event timings in ISO format, while the other is intended to be read by the graphical web schedule and has timings in UNIX format. The former is also used to display the schedule in tabular form on the website.}

\textcolor{black}{As described above, the data for airmass plots of potentially interesting targets are generated via a \texttt{Python} script; these data are then stored in their own worksheet to be read by the website. The information contained therein is not intended to be human readable, therefore the worksheet is not formatted and has no headers. Data from previous weeks are cleared once observations have taken place to make  the Google Sheet load quickly.}




\textcolor{black}{Once the data for the airmass plots have been generated in advance of a weekly meeting, the upcoming events can be viewed on the website on the target selection page. In order to facilitate fruitful discussions during the meeting, there is a box where team members can add their preferences and thoughts for upcoming scheduling. These thoughts are stored on the Google Sheet to be displayed in the appropriate place on the website.}


\subsubsection{Displaying the data: The website}
\label{sec:website}

\textcolor{black}{For effective collaborative working in an international team, it is essential that everyone can access all the information in a convenient way. In addition, Google Sheets are handy but can be easily altered mistakenly. With this in mind, we developed a website for the \astep team using \texttt{HTML, CSS} and \texttt{JavaScript}.}

\begin{figure*}[t]
    \centering
    \includegraphics[width = 0.9\textwidth]{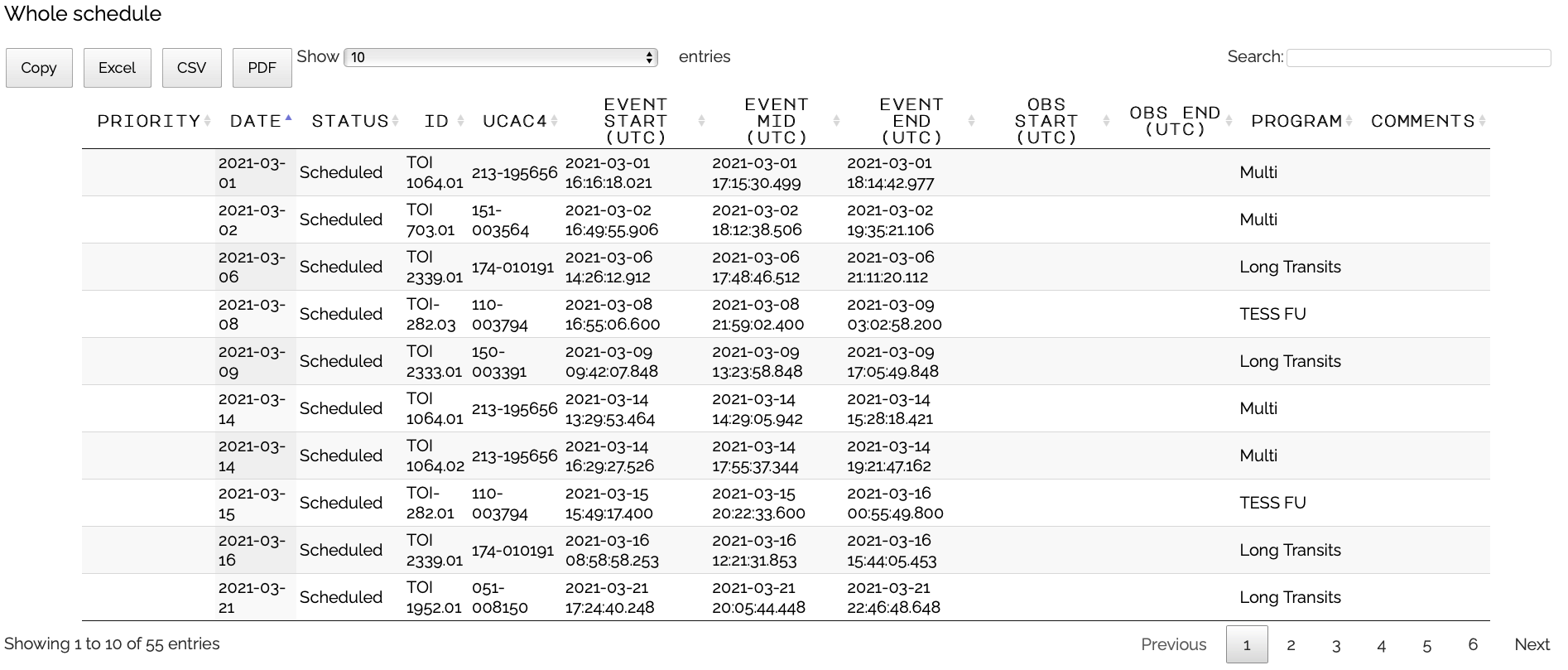}
    \caption{Screenshot of the DataTables display of the schedule as it appears on the ASTEP website.}
    \label{fig:dt_sched}
\end{figure*}

\textcolor{black}{The essential feature the website provides is access to data (see Fig.~\ref{fig:toolkit}), and this is done via DataTables\footnote{\url{https://datatables.net}}, a \texttt{JavaScript} package for interactive \texttt{HTML} tables. Information from each of the Google Sheets is fed in as \texttt{JSON}. As a result team members can access fully paginated, searchable and sortable \texttt{HTML} tables. These tables can also be exported in four formats; a screenshot of the schedule as displayed in a DataTable is presented in Fig.~\ref{fig:dt_sched}}.

\begin{figure*}[t!]
    \centering
    \includegraphics[width=0.8\textwidth]{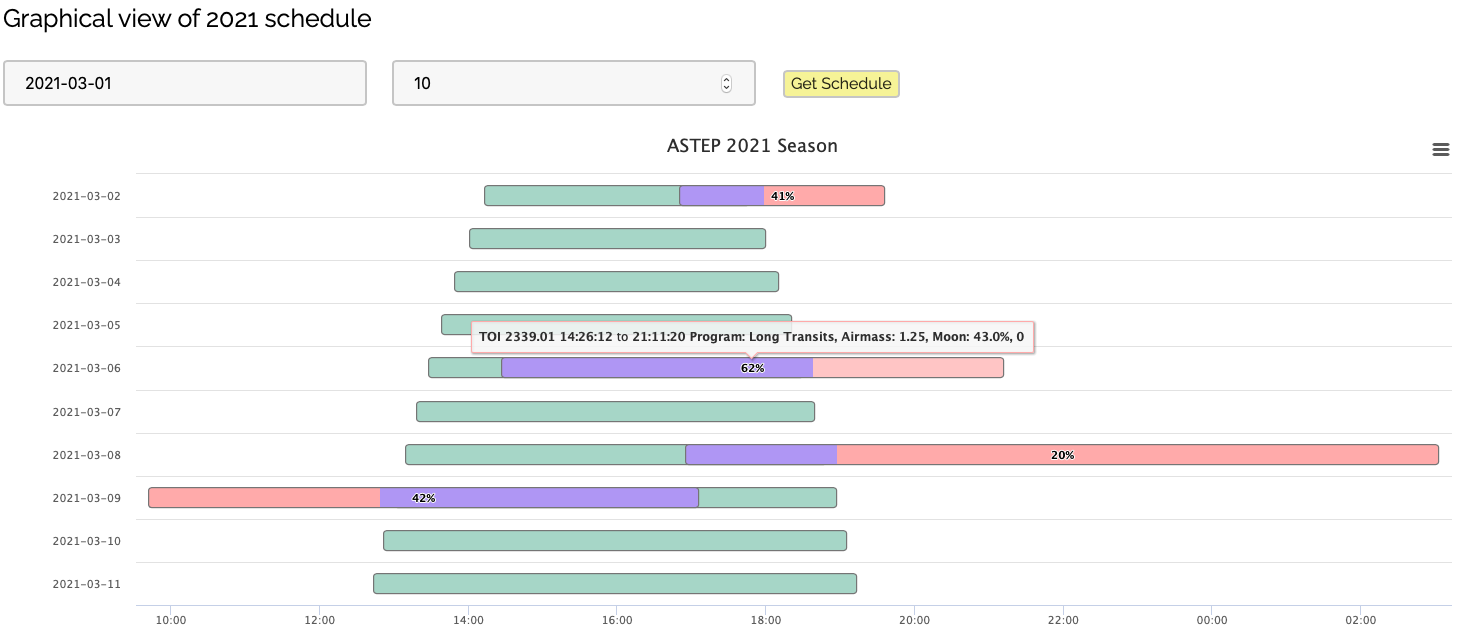}
    \caption{The graphical web schedule, showing user inputs and options. When the cursor is hovered over an observation bar, the tooltip shows the name of the target and the timing of the event, as well as the airmass and moon illumination during the event.}
    \label{fig:web_sched}
\end{figure*}

\textcolor{black}{In the hope of making the website experience as user-friendly as possible, we developed an interactive graphical view of the schedule using the \texttt{JavaScript} plotting package Highcharts\footnote{\url{https://www.highcharts.com}}. In Fig.~\ref{fig:web_sched} we present a screenshot of this schedule view. The green bars represent the available observing time each night, defined as the time between the start of the evening civilian twilight and the end of the morning civilian twilight, where the sun is at $6^{\circ}$ below the horizon. Scheduled events are shown as purple bars, which change to red for any portion of the event that is not during the night. Hovering over an event with the mouse displays the target name, the event timings, the observing program the target belongs to, the airmass at the time of mid-transit, and the percentage of moon illumination. All of these features allow for simple choices to be made when events clash.}

\textcolor{black}{As described in Sections \ref{sec:tools} and \ref{sec:gsheet}, we sought to incorporate all the most useful elements of the airmass plot \texttt{.pdf} files into our new system. To this end, we included the Target Selection page in the website, which displays the airmass plots for the top 30 scored transiting candidates from the \texttt{TTF}. We also see all the pre-scheduled events, allowing us to choose which observations allow us to make best use of the available time each night. Opportunistic observations can then be scheduled directly from this page and sent to the Google Sheet, removing any need for manual input. }


\begin{figure*}[t!]
    \centering
    \includegraphics[width=0.8\textwidth]{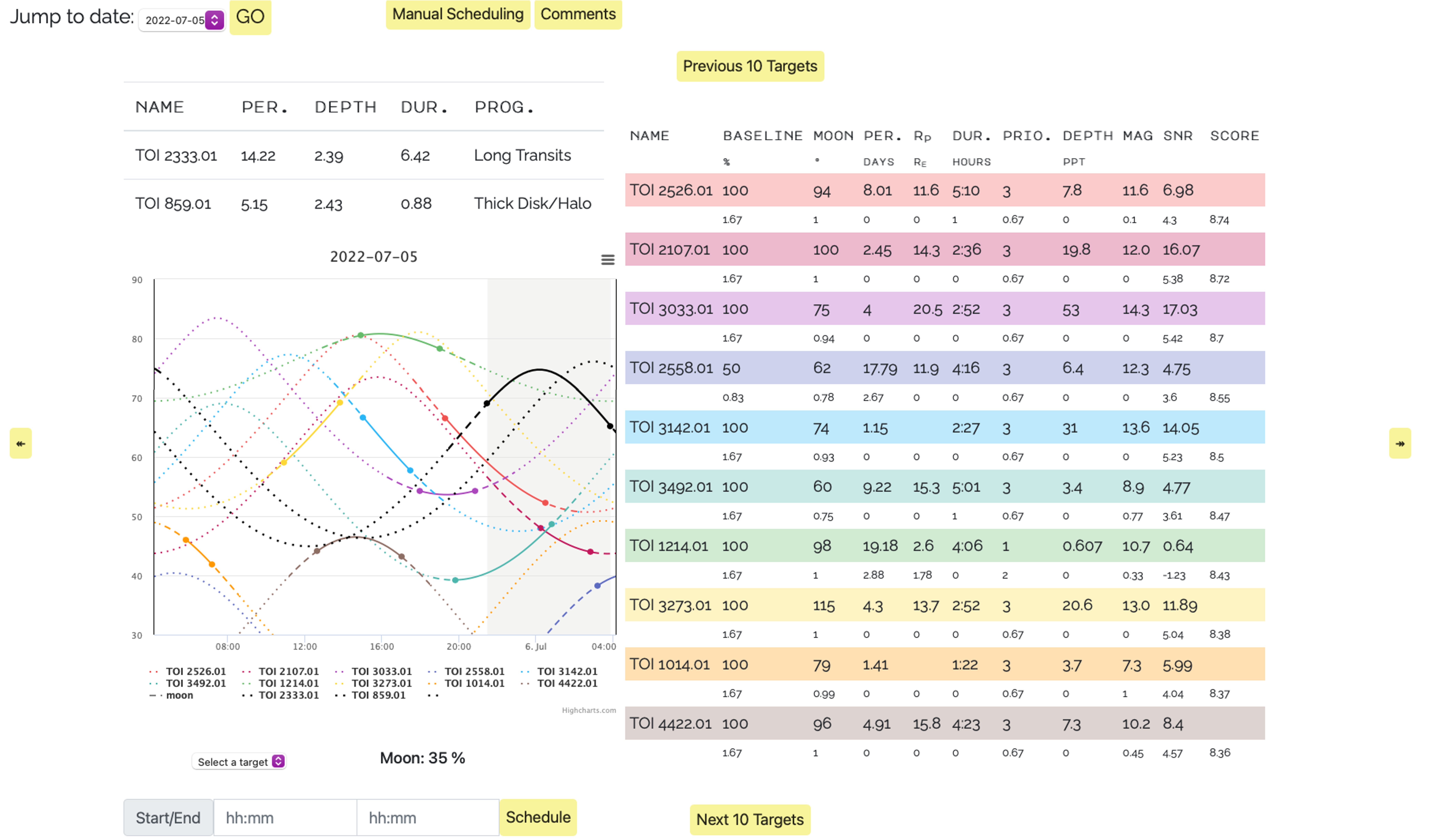}
    \caption{Screenshot of the \texttt{JavaScript} target selection toolkit as it appears on the website, with the layout adapted from Crouzet et al. (in prep). Interactive airmass plots are displayed on the left, and information for potential targets is shown on the right. Targets can be selected from the dropdown list and then scheduled with the `schedule' button.}
    \label{fig:airmass}
\end{figure*}


\subsection{Future Direction}

\textcolor{black}{During the 2021 observing season the version of the Google Sheets API we were using was unexpectedly retired, causing our entire toolkit to stop working. Upgrading to the new version of the API took approximately two weeks, during which time we had to revert to manual target selection. With this in mind, we aspire to phase out our use of G-Suite in future. This will be achieved by storing our data in databases as they can be easily written to and read by \texttt{Python} and web-based applications. The use of databases will therefore eliminate the need for a Google Sheet to store our data and the Sheets API to read/write information.}

\textcolor{black}{A further limitation which we will seek to address is that the full \texttt{Python} toolkit is locally stored and run on one team member's computer. This will be improved in future by using a \texttt{Python}-based web framework such as Django\footnote{\url{https://www.djangoproject.com}} or Flask\footnote{\url{https://flask.palletsprojects.com/en/2.1.x/}} to fully embed the toolkit in the website. This will allow all of the scripts to be run by anyone, from anywhere, at any time.}



\section{Data Analysis Pipeline}
\label{sec:pipeline}

\textcolor{black}{With our science goals shifting towards the competitive and fast-moving field of \textit{TESS} follow-up, the need for a fast automatic pipeline has grown. It's especially important for the pipeline data products to be as close as possible to the format needed for submission to ExoFOP in order to minimise the time spent in post-processing. Additionally, we want our data products to be versatile enough that we are able to extract additional information from our images without the need for full reprocessing, such as lightcurves of other stars detected in the field. Finally, as our team has grown to include collaborators from other astronomical sub-fields, there has been a growing need to produce one flexible data product that can be handled by others without the need for context-specific coding experience.}

\textcolor{black}{In this section, we will first describe in brief the \texttt{IDL} pipeline that has served \astep for the past six years, followed by a detailed description of the new \texttt{Python} pipeline in Section \ref{sec:prose}.}

\subsection{\texttt{IDL} Pipeline}

\textcolor{black}{The \texttt{IDL} data process pipeline, only briefly described here (see Abe et al. 2013\cite{Abe+2013} for a complete description) is a custom \texttt{IDL} code using classical aperture photometry routines from the well-known \texttt{IDL} astronomical library. Each science exposure frame is bias subtracted, dark corrected and the astrometric solution is computed using reference stars from the UCAC4 catalogue. Photometric lightcurves of about 1,000 stars are then performed through $10$ fixed circular apertures radii. The optimal calibrated lightcurve is then extracted using a set of
comparison stars.}
\textcolor{black}{This pipeline has served its purpose well and has allowed \astep to contribute to many publications since we joined TFOP in 2020\cite{Bouma+2020AJ,Dawson+2021,Burt+2021,Kaye+2022} plus lead our own discovery paper \cite{Dransfield+2022} with several others in preparation (Abe et al.in prep., Schmider et al. in prep., Triaud et al. in prep.). However, the cost of \texttt{IDL} and its declining usage means that maintenance of the pipeline is becoming highly specialised knowledge. Thus the need for a pipeline which is fast, modular and in a popular programming language: \texttt{Python}.}

\subsection{\texttt{Python} Pipeline}
\label{sec:prose}

\textcolor{black}{Rather than starting to develop the pipeline from scratch, we chose to build the pipeline around an astronomical data processing package written with \tess follow-up in mind: \prose{}.}

\textcolor{black}{\prose{}\footnote{\url{https://github.com/lgrcia/prose}} is an open-source \texttt{Python} package dedicated to astronomical image processing \cite{Garcia2022}. By featuring a wide range of pre-implemented processing blocks (from source detection to photometric extraction), it provides a framework to quickly assemble instrument-agnostic pipelines that are modular and easy to maintain. \prose{} is supplemented by convenient tools to manage and share the products of astronomical observations, such as the automatic generation of \tess follow-up reports, making it an ideal base for the new ASTEP+ pipeline. }

\textcolor{black}{In the following sections, describe the blocks and functionalities that form the core of the ASTEP+ pipeline, including custom modifications from the base \prose{} package. For a more comprehensive description of \prose{}, we direct the reader to the work where the package was first presented \cite{Garcia2022} and its online documentation\footnote{\url{https://lgrcia.github.io/prose-docs/build/html/index.html}}}









\subsubsection{Image Calibration}

\textcolor{black}{
We based ASTEP+'s pipeline on the \prose{} default photometric pipeline, which can be decomposed in two steps: The calibration and alignment of all images to produce a high-SNR stack image of the complete observation; And the extraction of the fluxes from the brightest stars in the field, using a wide range of size-varying apertures. The calibration sequence starts with the selection of a reference image, on which \textit{n} reference stars are detected, later used to align the rest of the science images on. Then, each raw image sequentially go through bias, dark and flat calibrations, before being trimmed for overscan pixels. For each image, the \textit{n} brightest stars are detected and compared to the reference ones in order to compute the affine transformation to the reference image. This is done using \texttt{twirl} \footnote{\url{https://github.com/lgrcia/twirl}} (a simplified \texttt{Python} implementation of \texttt{Astrometry.net}\cite{lang2010}). Finally, a stack image is created from the aligned images (transformed using bi-linear interpolation), while the unaligned calibrated images are saved to perform aperture photometry on.}

\subsubsection{Aperture Photometry}

\textcolor{black}{Two sequences are used in \prose{} to perform aperture photometry on the calibrated images. The first sequence sets the number of stars to be detected on the stack image to 1000. This ensures the detection of the target star even in crowded fields using the DAOFindStars block based on a \texttt{photutils}\footnote{\url{https://photutils.readthedocs.io/en/stable/index.html}} implementation of DAOPHOT. An elliptical two-dimensional Moffat model of the stack effective PSF is fitted in order to scale the photometric aperture radii. A total of forty circular apertures is used to extract the flux of all the detected stars in the following sequence. The position of the apertures on each calibrated image is calculated using the inverse transformation matrix computed in the calibration step. To correct for possible errors in these coordinates, the positions of the apertures are recomputed using the centroiding algorithm \texttt{Ballet}\footnote{\url{https://github.com/lgrcia/ballet}}, a convolutional neural network trained to predict accurately centroid positions. The flux is then extracted for each aperture and stored in a \texttt{.phot} file along with the values of systematic effects affecting the observation (position shifts on the detector, sky background, airmass and FWHM).}

\subsubsection{Plate Solving and Target Identification}

\textcolor{black}{\prose{} requires human intervention for the purpose of target selection on the stacked image; in order to automate this step we implement a local installation of \texttt{Astrometry.net}\cite{lang2010}. \texttt{Astrometry.net} is a robust and widely used system to provide blind plate solving of astronomical images. It works by extracting several patterns of four stars (asterisms) from query images, computing a hash code for each, and searching indexes for matching hash codes. The result is the  WCS (World Coordinate System) header information providing the pointing, scale and orientation of the query image.}

\textcolor{black}{Index files have been produced for many astronomical surveys in various colours. To ensure the best coverage, providing the best chance of an astrometric solution, we use the full set of \textit{Gaia} and \textit{2MASS} index files.\footnote{Available for download from \url{http://data.astrometry.net}}}

\textcolor{black}{Unlike the \texttt{IDL} pipeline, the new pipeline only requires that we plate solve the stack image as the field rotation between each image is calculated by the \texttt{twirl} block by \prose{} during the calibration stage of the pipeline.} 

\textcolor{black}{ASTEP+'s images are centered on the guide star, and its coordinates, together with the image pixel scale determined by the astrometric solution, are used to define a search cone. Under normal operation a call would be made to MAST (Barbara A. Mikulski Archive for Space Telescopes) via \texttt{astroquery}\cite{astroquery} to find all stars in this cone for a given catalog. In order to remove this need, we instead make use of a local version of the \textit{TESS} Input Catalog (TIC)\cite{TICv8}, saved in an \texttt{SQLite} database, to perform the search. The WCS header information is then used to transform celestial coordinates into pixel coordinates, and the closest aperture to the target is selected.}

\textcolor{black}{The decision to move toward using the latest version of the TIC as our main reference catalog instead of UCAC4\cite{UCAC4} was motivated by its completeness. Version 8 of the TIC includes all UCAC4 sources, as well as {\it Gaia} DR2 sources and several other large catalogs, making it the most complete catalog currently available.}



\subsubsection{Differential Photometry}

\textcolor{black}{Differential light curves are built from raw light curves using the algorithm presented in Broeg et al. 2005\cite{Broeg2005} (implemented in \texttt{Python} within \prose{}). This method consists of building an artificial comparison light curve using the weighted sum of all available stars in the field. The weight attributed to each star is computed through an iterative process that favours stars displaying lower variability and higher signal-to-noise ratio, more likely to feature systematic signals. The optimal aperture is then chosen as the one minimising the white noise estimated using the median standard deviation of points per five minute bins. All the light curves for each aperture are also stored in the \texttt{.phot} file allowing for a manual check if needed.  }


\subsubsection{Lightcurve Modelling}

\textcolor{black}{With the \texttt{IDL} pipeline, the biggest interaction required in post-processing has always been the modelling of lightcurves after they arrived in Europe. However, there are many simple cases where the detected transit matches the ephemerides well; to save time for these objects, we have written a new module for \prose: \texttt{auto\_modelling}.}

\textcolor{black}{All our lightcurves are correlated with airmass and background sky level. Additionally, variations in the FWHM (the full width at half maximum) can introduce spurious signals to the data. For this reason, all lightcurves are detrended by fitting an order two polynomial of time simultaneously to airmass, FWHM  and the background sky level. This is done at the same time as the transit modelling. There are sometimes additional signals in the data caused by motion of the telescope in the night ($\rm dx$ and $\rm dy$). Rather than detrend on these parameters, we reject any images where either of these parameters are greater than 5. }

\textcolor{black}{Transit fitting is implemented using \texttt{exoplanet},\cite{exoplanet} a flexible toolkit for modelling exoplanets using \texttt{PyMC3}\cite{pymc3} for MCMC (Markov Chain Monte Carlo) modelling. Priors on the transit parameters are taken from the image headers, while host star priors are drawn from the local TICv8 database. In cases where the stellar $\rm T_{eff}$, $\rm [Fe/H]$ and $\rm log g$ are available, quadratic limb darkening coefficients are also calculated using \texttt{PyLDTK}\cite{pyldtk}. These are then set as normal priors for the fit; where host data is not available, solar values of stellar mass and radius are used instead. In these cases uniform priors between $0-1$ are used for the limb darkening coefficients.}

\textcolor{black}{The automatic lightcurve modelling is not always successful. The most common reasons for failure are missing host star parameters, ambiguous detections and non-detections of transits, and transit timing variations (TTVs). In all these cases, manual post-processing is carried out after files have arrived on the server in Europe.}




\subsubsection{Delivery of Data Products}

\textcolor{black}{As described in Section \ref{sec:intro}, one of the biggest limitations that comes with an Antarctic telescope is the very limited internet connection. This is of course the motivation for our automatic data processing on-site, but it also limits how we access the data products output by the pipeline.}

\textcolor{black}{The data products produced by the pipeline fall into two categories: lightweight and heavyweight. Lightweight data products include a \texttt{.png} image of the target lightcurve and a \texttt{.csv} file containing the target flux and systematics. These products are emailed to all team members immediately after the pipeline finishes running to facilitate rapid inspection. In cases where the automatic lightcurve modelling and detrending has been successful, these two products are all that is needed.}

\textcolor{black}{The largest data products are the .phot files containing the aperture photometry for up to 1000 stars in the field as well as all the metadata for the observation, and the stack image produced during the reduction. These files are sent to a local server in Concordia using the \texttt{Python} package \texttt{paramiko}\footnote{https://www.paramiko.org}; this server has a folder which synchronises with a corresponding folder on a server in Rome. Depending on the size of the files, they take between $\rm 12-24\,hours$ to arrive in Europe; they are then available for download for any necessary post-processing.}



\subsection{Results}

\begin{figure}[t!]
    \centering
    \includegraphics[width=0.8\textwidth]{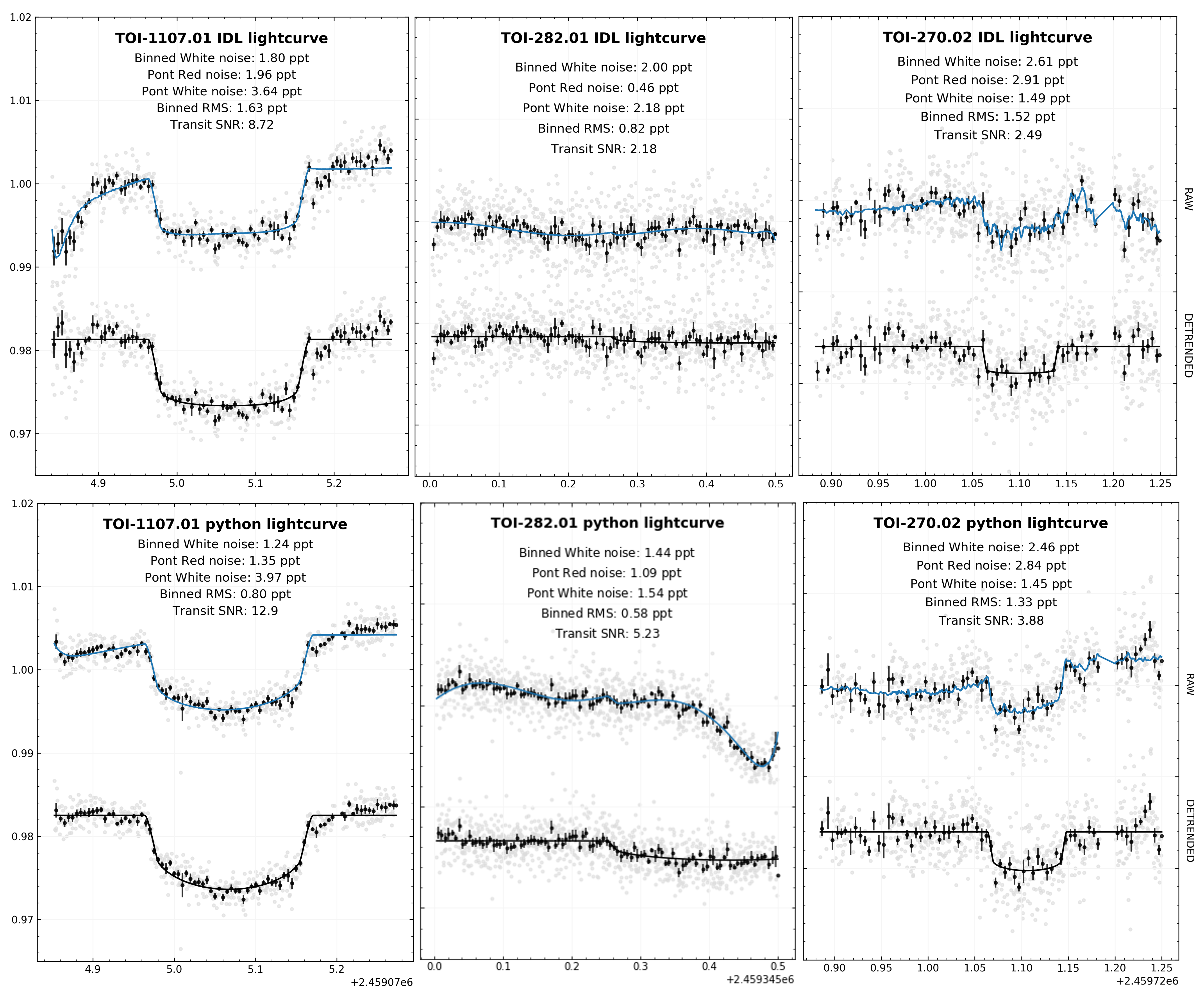}
    \caption{Comparison of lightcurves produced by the two pipelines. The upper panels show the lightcurves produced by the \texttt{IDL} pipeline, while the lower panels show the results of the \texttt{Python} pipeline. Grey points are normalised fluxes and black points are binned in 5 minute bins. The upper lightcurves are raw fluxes with the blue line indicating the systematics + transit model. The bottom lightcurves have had the systematics trend removed. Left panels: observation of TOI-1107.01 on 13 August 2020. Middle panels: observation of TOI-282.01 (now HD~28109~b) on 10 May 2021. Right panels: observation of TOI-270.02 (now TOI-270~d) on 21 May 2022.}
    \label{fig:lc_comps}
\end{figure}


In this section we compare the outcomes of the two pipelines, making quantitative comparisons where appropriate.


\subsubsection{Noise and SNR comparison}

\textcolor{black}{Following the method described in Garcia et al. (2022)\cite{Garcia2022}, we compare the automatic lightcurves produced by the \texttt{Python} and \texttt{IDL} pipelines using four metrics: the binned white noise, the white and red noise following Pont et al. (2006)\cite{Pont2006}, and the transit signal-to-noise ratio (SNR). All lightcurves used for the comparison were detrended by fitting a polynomial in time to the airmass and background sky level only; we then modelled each transit by fitting a simple Keplerian orbit using \texttt{exoplanet}. In all comparison cases, we placed uniform priors of $\mathcal{U}(0-1)$ on quadratic limb darkening coefficients. Using the resulting transit models, we also compare the binned root mean square (RMS) scatter of the lightcurves, calculated as the standard deviation of the binned residuals. The results of our comparison are presented in Fig.~\ref{fig:lc_comps}.}

The left-hand panels of Fig.~\ref{fig:lc_comps} show a deep ($\rm 9\,ppt$) transit observed during the 2020 season. Upon visual inspection, we see that there is less scatter in the \prose{} lightcurve than in the \texttt{IDL} plot, resulting in almost $\rm 1.5\, x$ higher SNR detection. We also see the effects of the stellar limb darkening more clearly in the transit shape of the \prose{} lightcurve.


In the middle two panels of Fig.~\ref{fig:lc_comps} we present a detection of the shallow ($\rm 0.85\,ppt$) transit of TOI-282.01 (now HD\,28109 b\cite{Dransfield+2022}) observed during the 2021 observing season. This system has large TTVs (transit timing variations) and the detected transit ingress was $\rm \sim72\, mins$ later than expected. We can see that while the \texttt{IDL}
pipeline does not clearly detect the transit, the \prose{} lightcurve has binned RMS scatter smaller than the transit depth, allowing for a conclusive detection of the event despite the large TTV.


\textcolor{black}{In the right-hand panels of Fig.~\ref{fig:lc_comps} we present the lightcurves of TOI-270.02, as observed in May of the current (2022) season. This planet is part of a three planet system where all planets find themselves in commensurate orbits leading once again to TTVs. As TTVs can be used to estimate planetary masses, ASTEP+ has a crucial role to play by measuring precise transit timings when systems are no longer visible from other southern observatories. Fig.~\ref{fig:lc_comps} shows that the new pipeline can yield more precise timings through higher SNR detections of transits.}


\subsubsection{Processing time}

\textcolor{black}{One of the most important considerations for science carried out in Antarctica is energy usage. Concordia generates electricity using two diesel generators with a third available in case of emergencies or malfunctions. Fuel for the generators reaches the station by traverse: a convoy of tractors pulling containers on skis, which crosses $\rm \sim1300\,km$ of ice over the course of $\rm 10-12$ days. 
Additionally, we must consider the greenhouse gas emissions resulting from computation. Each litre of diesel used by a generator emits $\rm 2.4-3.5$ kg of $\rm CO_2$\cite{diesel2012}. }

\textcolor{black}{On average, the \texttt{IDL} pipeline takes approximately $\rm 6$ seconds per image to run in full, while the new \texttt{Python} pipeline takes $\rm 1.6$ seconds. The addition of a second camera means that the pipeline will run for approximately twice as long, depending on the exposure times of the images in the respective colours. Even running the pipeline in full twice, our server will now spend half the time running intensive processes per day, therefore increasing CPU idle time. On average, power usage is increased threefold during computationally intensive processes when compared with idle time\footnote{\url{https://wccftech.com/review/intel-core-i9-9900k-8-core-cpu-z390-aorus-master-review/9/?beta=1}}, and with it the carbon footprint of the pipeline. }




\subsection{Limitations \& Future Work}

\textcolor{black}{The biggest limitaiton we currently face with the \prose{} pipeline is that we cannot yet process very defocused observations where the PSF (point spread function) is donut shaped. This does not present a huge problem in the current season as we have paused our observations of $\beta$~Pictoris, but should we return to observing this very bright target it will become necessary to defocus the telescope once again. When the pipeline is updated during the next summer service mission, we intend to test new defocused PSF modelling blocks in order to ensure these observations can be processed in future.}



\section{ASTEP+ in the Future}
\label{sec:future}


At the moment ASTEP+ is mainly pursuing the validation of \tess transiting exoplanet candidates. Most of the easy pickings have been detected, and \tess will increasingly produce long period planets, candidates on fainter stars, or small planets producing weaker events. Whilst ASTEP+ is well geared towards the validation of longer orbital periods, there will be few targets and events are few and far in-between. This means that the ASTEP+ project will need to evolve. We are currently investigating photometric monitoring in coordination with X-ray satellites such as XMM, Chandra, AstroSat, for the study of flaring stars\cite{Lalitha2020}. Of course, the monitoring of known transiting exoplanets will have to continue in order to refine the ephemerides and plan observations by programs such as JWST and Ariel as efficiently as possible \cite{Kokori+2022}. 

Another activity we are keen on pursuing is to repeatedly detect the transits of planetary systems where at least two planets have orbital periods near an integer ratio. This configuration means their gravitational interaction are detectable (producing Transit Timing Variations, TTVs), and would allow us to measure the planet's masses such as done already in papers where ASTEP observations were crucial \cite{Dransfield+2022, Kaye+2022}.  In addition, ASTEP+ has produced the first ground-based detection of a circumbinary planet transit (Triaud et al. in prep). To support this science, measuring Eclipse Timing Variations (ETVs) produced by the eclipsing binary stars at centre of such system is also a source of information about the planets. In both science cases, all transits/eclipses are useful. Our unique position near the pole and opposite from Chile means we can collect a high yield of events using similar arguments to those described in Section~\ref{sec:long}.



In addition, we are currently developing a new dedicated direct-drive mount adapted to Antarctic conditions. This should maximize observation efficiency and pointing stability, thereby producing higher quality lightcurves. With this mount, we intend to check whether some transients are detectable, and test a rapid response mode to detect the electromagnetic counterparts of Gravitational Wave events, or be used has to follow up observations where a transient detected by the Vera Rubin telescope (in Chile) has set as seen from their location, but is still visible from our location thanks to our proximity to the South Pole. We are also investigating the possibility to search for and detect interstellar asteroids within the frames obtained during exoplanet transit observations. The idea is to leverage the fact that our field of view is mainly out of the ecliptic. Anything that moves is less likely to be from the Solar system.

The highest potential of Astronomy at Concordia however lies with observations in the infrared, particularly in the K band. As for the high plateaus of Antarctica in general, Concordia is characterized by a high sky transparency, low water content and low thermal background that makes it one of the best sites on Earth for infrared Astronomy \cite{Burton+2016}. Combining observations in the visible with ASTEP+ to observations in the infrared with a new telescope of a similar aperture could provide a real breakthrough for the monitoring of exoplanetary atmospheres and exoplanets around low-mass stars and for the study of counterparts of gravitational wave events. 

\section{Conclusions}
\label{sec:conclusions}

\textcolor{black}{In this work we have presented a suite of tools developed for scheduling time-domain astronomical observations. While these systems have been written with ASTEP+ in mind, they are easily adaptable for other observatories focused on observations with strong time constraints.}

\textcolor{black}{We have also presented a new automatic data analysis pipeline written in \texttt{Python} and built around \prose{}. The new pipeline produces lightcurves that on average have lower red and white noise, lower scatter, and therefore allow for transits to be detected with higher SNR compared with the system it replaces. Additionally, as the pipeline runs significantly faster we will also decrease our energy usage and carbon footprint resulting from computationally intensive reductions. This remains true even taking into account the increase in data generated by having two cameras instead of one.}

\textcolor{black}{Finally, we have outlined the future directions for the ASTEP+ project, including new collaborations and synergies at the forefront of time-domain astronomy. }


\acknowledgments 
 
This research is in part funded by the European Union's Horizon 2020 research and innovation programme (grants agreements n$^{\circ}$ 803193/BEBOP), and from the Science and Technology Facilities Council (STFC; grant n$^\circ$ ST/S00193X/1).
We acknowledge support from the European Space Agency (ESA) through the Science Faculty of the European Space Research and Technology Centre (ESTEC). 
ASTEP and ASTEP+ have benefited from the support of the French and Italian polar agencies IPEV and PNRA, and from INSU, ESA through the Science Faculty of the European Space Research and Technology Centre (ESTEC), the University of Birmingham, the laboratoire Lagrange (CNRS UMR 7293) and the Universit\'e C\^ote d'Azur through Idex UCAJEDI (ANR-15-IDEX-01). 
This publication benefits from the support of the French Community of Belgium in the context of the FRIA Doctoral Grant awarded to Mathilde Timmermans and Lionel J. Garcia.
MNG acknowledges support from the European Space Agency (ESA) as an ESA Research Fellow.
\bibliography{report} 
\bibliographystyle{spiebib} 

\end{document}